\documentclass{article}

\usepackage{amsmath}
\usepackage{PRIMEarxiv}
\usepackage[utf8]{inputenc} % allow utf-8 input
\usepackage[T1]{fontenc}    % use 8-bit T1 fonts
\usepackage{soul}
\usepackage{float}
\usepackage{booktabs}
\usepackage{pict2e}
\usepackage{multirow}
\usepackage{rotating}
\usepackage{amssymb}
\usepackage{hyperref}       % hyperlinks
\usepackage{url}            % simple URL typesetting
\usepackage{amsfonts}       % blackboard math symbols
\usepackage{nicefrac}       % compact symbols for 1/2, etc.
\usepackage{microtype}      % microtypography
\usepackage{graphicx}

\title{Comparative analysis of Realistic EMF Exposure Estimation from Low Density Sensor Network by Finite \& Infinite Neural Networks
\thanks{\textit{{
This work has been accepted for publication. Copyright may be transferred without notice, after which this version may no longer be accessible. Some changes are expected in the published version.
Mohammed Mallik is with INSA Lyon, Inria,  CITI, UR3720, France. (email:mohammed.mallik@insa-lyon.fr). Laurent Clavier is with IMT Nord Europe, France (e-mail: laurent.clavier@imt-nord-europe.fr)
Davy P. Gaillot is with Univ. Lille, CNRS, UMR 8520 - IEMN, F-59000, Lille, France. (email: davy.gaillot@univ-lille.fr)}}
}}

\author{%\Large
{Mohammed Mallik\textsuperscript{1}, Laurent Clavier\textsuperscript{2}, 
and Davy P. gaillot\textsuperscript{3}}\small}

\begin{document}

\maketitle

\begin{abstract}
Understanding the spatial and temporal patterns of environmental exposure to radio-frequency electromagnetic fields (RF-EMF) is essential for conducting risk assessments. These assessments aim to explore potential connections between RF-EMF exposure and its effects on human health, as well as on wildlife and plant life. Existing research has used different machine learning tools for EMF exposure estimation; however, a comparative analysis of these techniques is required to better understand their performance for real-world datasets. In this work, we present both finite and infinite-width convolutional network-based methods to estimate and assess EMF exposure levels from 70 real-world sensors in Lille, France. A comparative analysis has been conducted to analyze the performance of the methods' execution time and estimation accuracy.  To improve estimation accuracy for higher-resolution grids, we utilized a preconditioned gradient descent method for kernel estimation. Root Mean Square Error (RMSE) is used as the evaluation criterion for comparing the performance of these deep learning models.

\end{abstract}

% keywords can be removed
% \keywords{First keyword \and Second keyword \and More}

\section{Introduction}

Wireless communication systems have seamlessly woven into our daily lives, becoming an essential part of modern existence. As a result, monitoring the phenomena associated with these systems, particularly radio-frequency electromagnetic field (RF-EMF) exposure, is of significant importance. 

Mobile devices and base stations that emit EMFs for radio communication are required to comply with regulatory guidelines for human exposure \cite{international2020guidelines, bailey2019synopsis} and it is crucial to monitor the potential effects of exposure to these systems. However, despite progress in this area, the complexity of monitoring EMF exposure remains high due to the wide range of influencing factors—such as materials, sensor configurations, and environmental topology. This complexity often necessitates extensive and time-consuming measurements, making the process costly and labor-intensive.

In our previous works, \cite{mallik2022towards,mallik2022eme, mallik2023eme, 10478001, 9569470},  EMF exposure was inferred by finite and infinite width convolutional neural models in indoor and urban environments, specifically in \cite{mallik2022towards,mallik2022eme,mallik2023eme}, reference full maps are used for training, where the datasets were generated from ray-tracing simulators \cite{egea2019vehicular,amiot2013pylayers}. In all these works, we raised a major concern:
\par 1. The generated datasets are imperfect, as they overlook several critical factors such as transmit power, antenna configurations, other EMF sources, traffic conditions, and weather influences like rain, etc. Additionally, creating these datasets demands both time and expertise. Moreover, often ML models perform poor on real world data than synthetic data \cite{ALKHALIFAH2022101}. 

In this work, we want to investigate how well our models \cite{10478001} and \cite{9569470} can generalize EMF exposure from a real-world sensor network which does not require a large training set or reference full maps for training. Given that infinite-width neural networks are relatively new in the context of exposure estimation \cite{10478001}, though evaluated on other tasks \cite{jacot2018neural, arora2019exact, pedregosa2011scikit}, our goal is to evaluate how effectively they perform when applied to real-world datasets. In both methods, we select a \textit{prior} to reconstruct EMF exposure accurately, named as Local Image Prior (LIP). We compare our results to find which methods performance is the best. 

The paper is organized as follows. In Section \ref{s1}, we describe the methodology that includes EMF exposure reconstruction, problem formulation, dataset, methods, and model details. Experimental results are presented in Section \ref{s3}. The conclusion is given in Section \ref{s4}.

\section{Methodology} 
\label{s1}
\subsection{Inverse problem formulation}
This work focuses on reconstructing the RF-EMF exposure map for a 1 $km^2$ rectangular region, discretized into an $M \times N$ grid, located in Lille, France, specifically covering the Wazemmes and Euratechnologies districts. The reconstruction is based solely on data from sparsely distributed, randomly positioned fixed sensors within the region. Each sensor, positioned at the coordinates $(m,n)$ on the grid, where $m \in \{1,\dots,M\}$ and $n \in \{1,\dots,N\}$, is able to measure local exposure levels $e_{(m,n)} \in \mathbb{R}^S$, with $S$ representing the number of sensors.  The challenge of estimating exposure $e \in \mathbb{R}^{M\times N}$ at every grid point is framed as an inverse problem. The goal of EMF exposure reconstruction is to find a function $f_\theta:\mathbb{R}^S\rightarrow \mathbb{R}^{M \times N}$ that can predict exposure at locations where no sensors are present, with $\theta$ representing the function's parameters. In this work, a neural network model $f$ is employed to perform this reconstruction, incorporating models such as EME-CNTK \cite{10478001} and GLIP  (Generative Local Image Prior) \cite{9569470}.
% \section{Proposed Method}
% \label{s2}
% In this section, we present the proposed method for
% performing exposure reconstruction with a deep convolutional neural
% network, and detailing the proposed architecture.

\subsection{EME-CNTK}

In \cite{wang2020sensor}, the neural network function used to model the exposure in Paris trained with Adam optimizer and Mean Squared Error (MSE) loss is given by:
\begin{equation}\label{eq:1}
\Psi _{\psi}(x) = \Psi_{\psi_{k}}\phi(\Psi _{\psi_{k-1}}\phi(\Psi _{\psi_{k-2}}\phi(...(\Psi_{\psi_1}(x))...)), 
\end{equation}
where $x$ is the input and $\Psi_{\psi_k}$ $\in\{1,\cdots,k\}$, are the layer functions followed by a non-linear function as activation $\phi$.
In \cite{jacot2018neural}, Jacot \textit{et al.} showed that an overparameterized or infinitely wide neural network trained under certain condition such as MSE loss and Stochastic Gradient Descent (SGD) optimizer, can be characterized by a continuous kernel function called the Neural Tangent Kernel (NTK). This is analogous to kernel regression using the NTK \cite{arora2019exact}. 

In this work, we train an infinitely wide convolutional neural network to compute the Convolutional NTK (CNTK) with a different number of layers, kernel size, and activation. The connection between the CNN model and the CNTK is in \cite{arora2019exact}:
\textit{Let $f(w): \mathbb{R}^P \rightarrow \mathbb{R}$ denote a neural network with initial parameters $w^{(0)}$. The \textbf{CNTK,} $K: \mathbb{R^d}\times \mathbb{R^d}\rightarrow\mathbb{R}$ is a positive semi-definite function given by:} 
\begin{equation}
K(x, x') = \langle \frac{\delta f(\theta, x)}{\delta w_{i,j,k,l}} \times \frac{\delta f(\theta,x')}{\delta w_{i,j,k,l}}\rangle,
\end{equation}
where $x$ and $x'$ are input samples, $\theta$ represents parameters of the network, $w_{i,j,k,l}$ denotes the summation over the indices of the convolutional filters and their corresponding weights $w^{(0)}$, and the $L$-th layer CNTK kernel is given by:

\begin{equation}
    K(x,x') =[\Theta^{L}(A,A')]_{i,j,i',j'}
\end{equation}
$A$ is the \textit{prior} - LIP, Which is a matrix encoding the structure of the coordinates labels.

% it is defined as ${A\in \mathbb{R}^{C\times H \times W}}$, where $H,W,C$ is the height, width and channels of the image.
\subsubsection{EME-CNTK+EigenPro}
In \cite{10478001}, EMF exposure is estimated in low resolution grids with size $32\times32$ and $64\times64$ by exact CNTK. This is a suitable option for low resolution grids, but infeasible for larger grids like $128\times128$ which affect the performance and accuracy of the estimation by exact CNTK. In this work, we employed a pre-conditioned Gradient Descent (GD) method EigenPro \cite{ma2017divingshallows,meanti2020kernel}, which efficiently scales the CNTK by training for higher resolution tensors. The iterative update rule with EigenPro preconditioning is:
\begin{equation}
    \alpha^{(t+1)} = \alpha^{(t)} - \eta \mathbf{P} (\mathbf{K} \alpha^{(t)} - \mathbf{y}),
\end{equation}
where $\alpha^{(t+1)}$ are the coefficients for the kernel expansion at iteration $t$, $\eta$ is the learning rate, $\mathbf{K} \alpha^{(t)}$ represents the kernel-based predictions and $\mathbf{y}$ is the vector of target values. Finally, $\mathbf{P}$ is the EigenPro preconditioner, designed to accelerate convergence by adjusting the gradient based on the top eigenvalues of the kernel matrix $\mathbf{K}$.
Once the coefficients $\alpha_i$ have been learned, the prediction for a new input $A_i$ is given by:
\begin{equation}
      \Upsilon''_i = \sum_{i=1}^n \alpha_i k(A_i, A_i'),
\end{equation}
\subsection{GLIP}

The exposure map image is reconstructed by a deep generative network function $f$ in \eqref{eq:1} where $A$ is the \textit{prior} - LIP of the network. In this case, the objective function in \cite{9569470} $E(f(\theta|A); A)$ is only calculated from the observed values in $A$. Hence, a binary mask $m \in \{0,1\}^{H\times W}$ is used to take into account only the points observed in the sparse exposure image $\Psi$. $\Psi_i$ denotes the reconstructed map at iteration $i$ during the training. hence, the objective function becomes: % in \eqref{eq:4}: 
\begin{equation}\label{eq:glip}
    E(A,A_i) = ||(A-A_i) \odot m||^2.
\end{equation}
The norm %within this expression 
in \eqref{eq:glip} will be the squared error between the observed and predicted values. 
The $\odot$ represents the Hadamard product or element-by-element multiplication.

\subsection{Model details}
For EME-CNTK and GLIP, we kept the same configuration as described in \cite{10478001,7338410}. 
% For Noisy-EMF, the $U$ and $N$ model has same architecture which was taken from \cite{hwang2023autoencoder} and The input and output dimensions are 16384. 
All experimented model's details are given in Table \ref{tab:table3}:
\begin{table}[!htp]
\caption{Model parameters.}
\centering
%\resizebox{\textwidth}{!}{
\begin{tabular}{|l|c|c|c|}
\hline
\multicolumn{1}{|c|}{Parameters} & \multicolumn{1}{c|}{EME-CNTK+EigenPro} & \multicolumn{1}{c|}{GLIP} \\ \hline
Optimizer & Gradient Descent & Adam \\\hline
Non-Linearity & LeakyReLU & LeakyReLU \\ \hline
% Upsampling &  Nearest Neighbour &  Nearest Neighbour & Nearest Neighbour   \\\hline
% Downsampling                               &  Strided convolution2D   \\\hline
Loss function & MSE & MSE\\\hline
Epochs & 350& 150\\\hline 
\end{tabular}
\label{tab:table3}
\end{table} 

\section{Results}\label{s3}

To evaluate our methods, two scenarios are considered. The performance is investigated on 4 and 2 sensors which are not considered in the learning phase in two 1 km$^2$ areas. We tested the EME-CNTK and GLIP on the same test dataset with two priors, namely LIP and random normal prior (RNP).

\subsection{Evaluation metrics}
To evaluate the performance of our systems, the root mean square error (RMSE) was used:
\begin{equation}
\text{RMSE} = \sqrt{\frac{1}{n} \sum_{i=1}^{n} \left(y_i - \hat{y}_i\right)^2},
\end{equation}
where $(y_i -\hat{y_i})$ denotes the error between the reference value $y_i$ and the predicted value $\hat{y_i}$ and $n$ is the number of points in the image.

% We also use the mean absolute error (MAE), given by:
% \begin{equation}
%     \mathrm {MAE} =\frac{1}{n}\sum _{i=1}^{n}\left|Y_{i}-\hat{Y}_{i}\right|,
% \end{equation}
% which limits the impact of large errors on the resulting error metric.

\subsection{Implementation Details and Dataset}
\begin{figure}[h]
\centering
\includegraphics[scale=0.3]{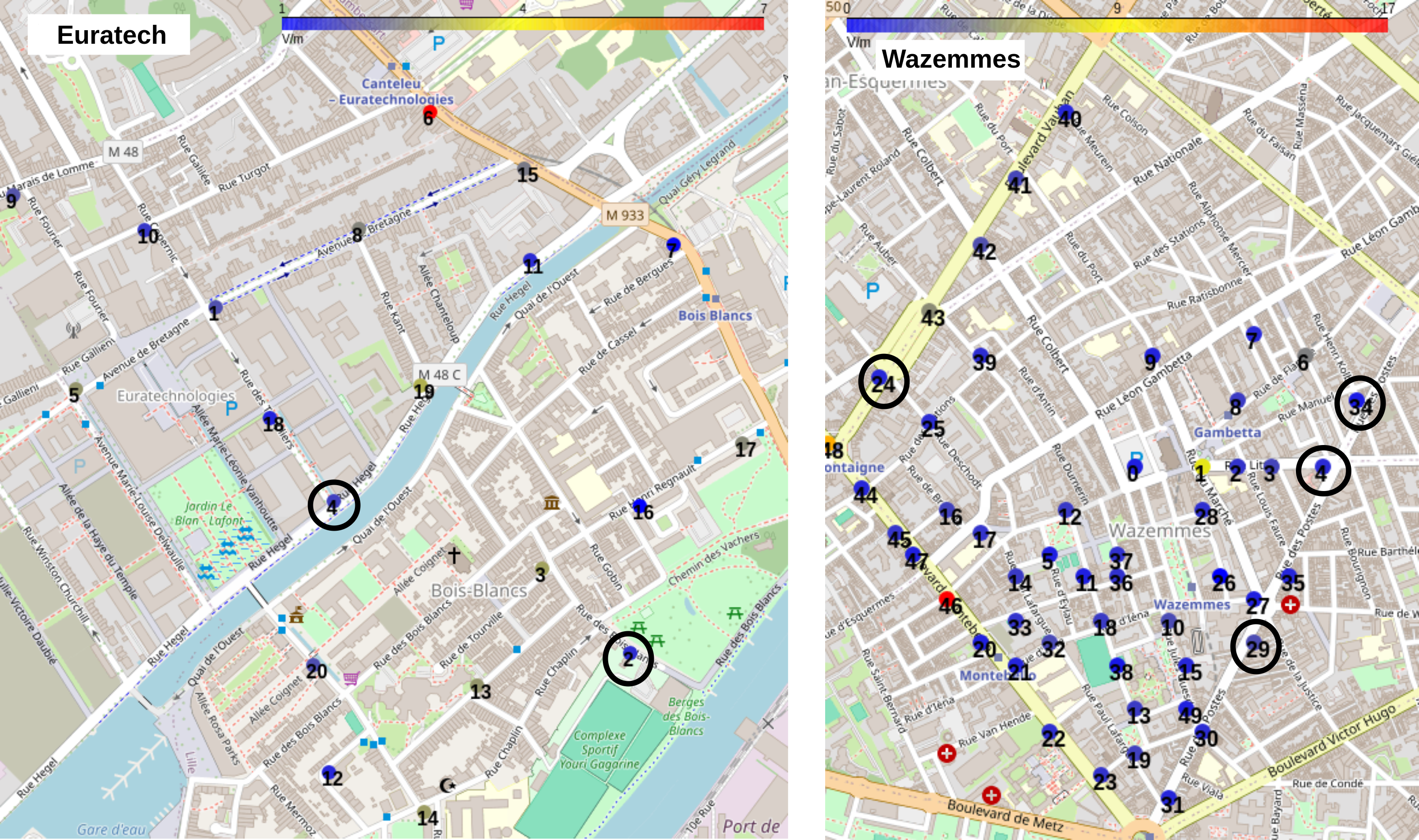}
\caption{Visualization of the region of interest. This showcases the experimental 1$km^2$ areas in Wazemmes and Euratechnologies, where black circles show the sensors kept for error calculation: 2 and 4 in Euratechnologies and 4, 24, 29 and 34 in Wazemmes, respectively.}
\label{fig:warawaz}
\end{figure}
The region of interest (RoI) for this study encompasses two 1 km² areas within the city of Lille, specifically the Wazemmes and Euratechnologies districts. The sensor network, maintained by Métropole Européenne de Lille (MEL), consists of 50 and 20 sensors, respectively. This data was obtained using a web service that allows queries filtered by date, exposure level, humidity, etc. every 2 hours. Dataset consists of 100 hours from 23 November 2023 in csv format. Data has been cleaned to avoid repeated entries, outliers, NaN values and normalized. To evaluate reconstruction accuracy, 4 and 2 sensors in Wazemmes and Euratechnologies were reserved for error calculation and not used in the reconstruction process. A visual representation of the RoIs and the sensors used for evaluation is depicted in Fig.~\ref{fig:warawaz}.

Moreover, the environment (roads, buildings, etc.) has been taken from OpenStreetMap to incorporate the city topological effect by suppressing the building locations in the predicted maps $\Upsilon''$. For all methods, the dataset consists of images, and each picture is $A\in \mathbb{R}^{1\times 128\times128}$. 
% We tried to follow Pytorch notations for all methods, as far as possible. 

\subsection{Visual Analysis}
\subsubsection{EME-CNTK \& GLIP}

\begin{figure*}[h]
    % \centering
\includegraphics[width=\textwidth]{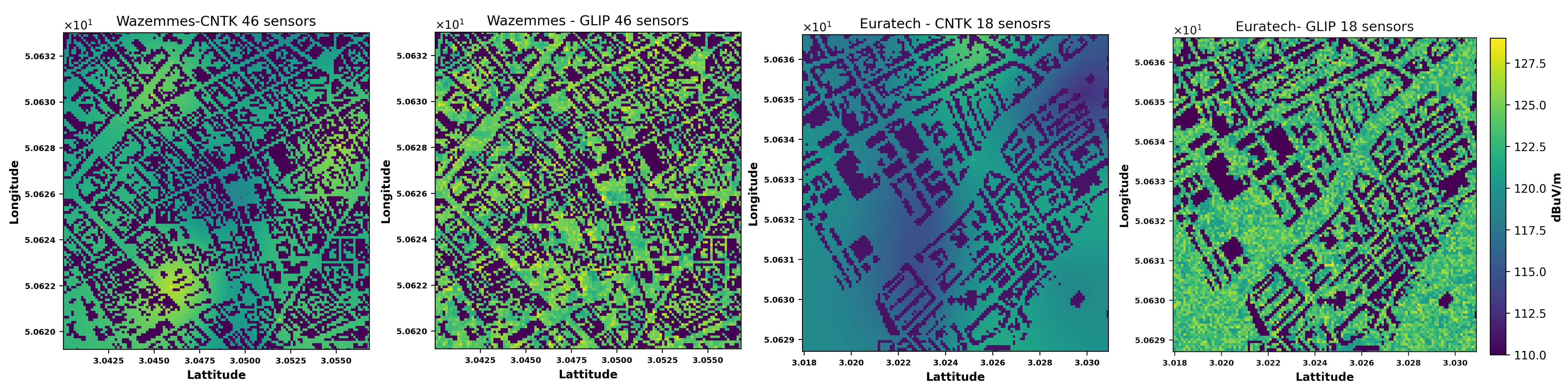}
\caption{Experimental results for EME-CNTK$_{EigenPro}$ \& GLIP by 46 and 18 sensors in Wazemmes and Euratechnologies districts, respectively.}\label{fig:res}
\end{figure*}

The reconstructed maps generated by the EME-CNTK$_{EigenPro}$ and GLIP methods are shown in Fig.~\ref{fig:res}. It is evident that GLIP struggles to generalize and accurately reconstruct the exposure in the RoIs. Previous work in \cite{9569470} demonstrated a similar limitation, where GLIP failed to reconstruct exposure in a $32 \times 32$ image with only 20 sensors. Likewise, in the current scenarios, with just 46 and 18 measurement points available, GLIP falls short and fails to estimate the exposure over a large $128 \times 128$ image in 1 $km^2$ area. GLIP optimizes a neural network to generate exposure image based on the observed data. However, this approach tends to work better when a larger portion of the measurement points are available, allowing the network to generalize from known pixels to infer missing ones. In this case, the reconstruction error is notably high at $6.01 \times 10^{-1}$ V/m, particularly due to the lack of data points or sensors in specific areas, resulting in significant inaccuracies in the reconstruction.

A similar trend is observed with the EME-CNTK$_{exact}$, which also produces a high error of $8.59 \times 10^{-1}$ V/m. This is because using an exact CNTK is not practical for matrix imputation on a high-resolution grid, such as $128 \times 128$, further contributing to the reconstruction challenges in this scenario.
In contrast, EME-CNTK$_{EigenPro}$ demonstrates the ability to accurately estimate exposure on a $128\times 128$ grid in Wazemmmes district, achieving a low error of $1.99\times10^{-1}$ $V/m$. 
%It is to be noted that the maximum exposure was 6.0 $V/m$ in the dataset. [I THINK WE SHOULD REMOVE THIS SENTENCE.] 
Furthermore, the reconstructed results for both RoIs using EME-CNTK$_{EigenPro}$ are visually smoother compared to those produced by the GLIP method.

% \subsection{Evaluation protocol}

\subsection{Quantitative Analysis}
%\subsubsection{Impact of Sensor Density}
We present in Table \ref{tab:table_mse1} the RMSE in $V/m$ of the predicted and actual values of exposure for two sets of reference sensors over 100 images. 
\begin{table}[htbp]  
\centering
\caption{RMSE of the estimated exposure in {V/m}}\label{tab:table_mae}
% \captionsetup{position=below}
\label{tab:table_mae}
\begin{tabular}{ccccc} 
\toprule
RoI & CNTK & CNTK &CNTK & GLIP \\ & $exact$ & $EigenPro$& $EigenPro$&\\
&LIP &LIP & RNP&LIP\\
\midrule
W  & 8.59.$10^{-1}$ & 1.99.$10^{-1}$ & 8.59.$10^{-1}$& 4.96.$10^{-1}$ \\ 
E & 7.46.$10^{-1}$ & 4.59.$10^{-1}$& 8.70.$10^{-1}$&6.61.$10^{-1}$\\
\bottomrule
\end{tabular}
\label{tab:table_mse1}
\end{table}

As expected, as shown in Table \ref{tab:table_mse1}, training the CNTK with EigenPro (the CNTK was trained during 350 epochs) reduces the RMSE, leading to better reconstruction performance for both Wazemmes and Euratechnologies areas. 
However, when using exact CNTK, the RMSE is significantly higher than EME-CNTK$_{EigenPro}$ (for Wazemmes 46 sensors RMSE: 1.99.$10^{-1}$ and Euratechnologies 18 sensors RMSE: 4.59.$10^{-1}$ $V/m$). The \(4.59 \times 10^{-1} \, V/m\) error in Euratechnologies is slightly greater than that observed in the Wazemmes district. This indicates that utilizing just 18 sensors is insufficient for accurately estimating exposure across a 1 km² area within a larger grid. Similarly, the GLIP method errors are significantly higher (for Wazemmes 46 sensors RMSE: 4.96.$10^{-1}$ and Euratechnologies 18 sensors RMSE: 6.61.$10^{-1}$ $V/m$)  proving that the density of the sensor network is too low for a better generalization performance on a higher resolution grid like $128 \times 128$. Moreover, we also observe that, using prior $RNP$, the error is significantly higher than $LIP$ reconstruction (for Wazemmes 46 sensors RMSE: 8.59.$10^{-1}$ and Euratechnologies 18 sensors RMSE: 8.70.$10^{-1}$ $V/m$). 
\begin{figure*}[h]
    % \centering

\includegraphics[width=\textwidth]{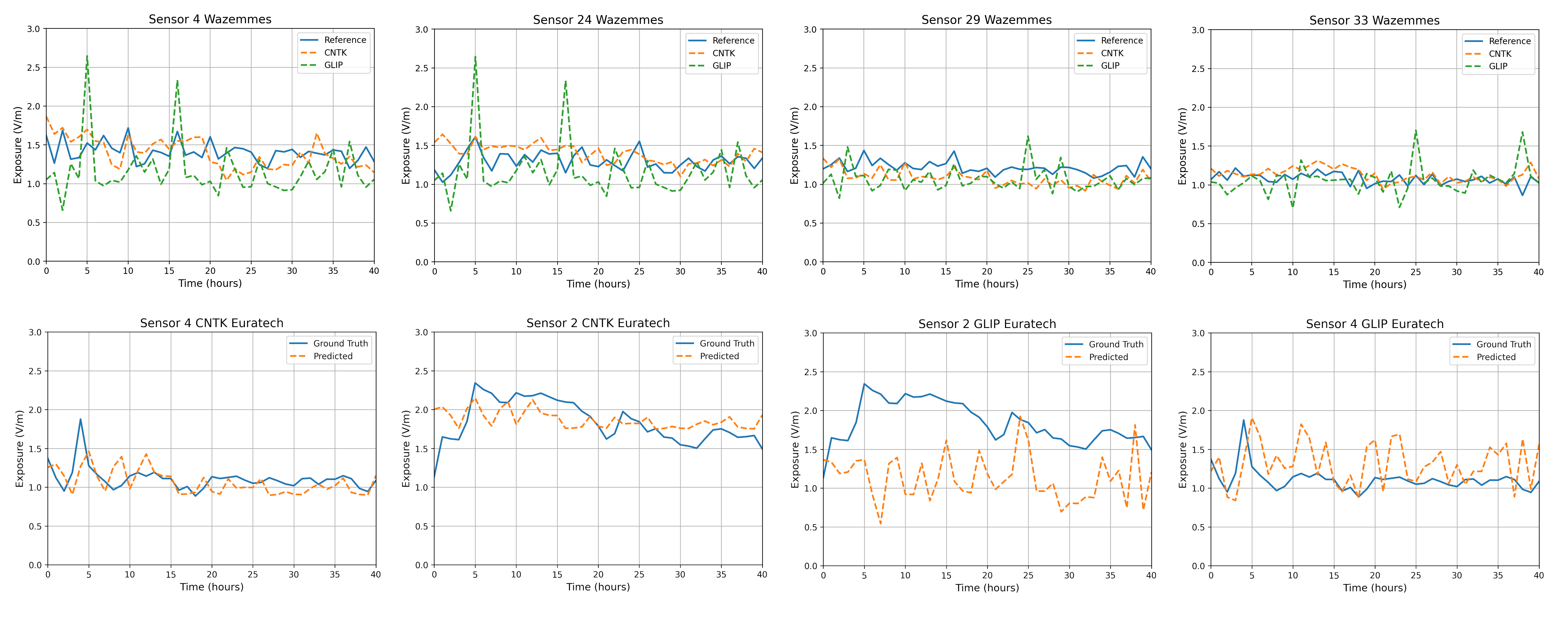}
\caption{The results of the CNTK and GLIP models showcases the reference and estimated EMF exposure levels. The 1st row presents the graphs of CNTK and GLIP for both reference and estimated exposure for sensors 4, 24, 29, and 33 in the Wazemmes district. In the 2nd row, the graphs depict the reference and estimated exposure for sensors 2 and 4 in the Euratechnologies district, with the first two graphs corresponding to CNTK and the latter two to GLIP. All results are shown when LIP prior is used.}\label{fig:evolution}
\end{figure*}
%It is to be noted that the maximum exposure level was  6.0 $V/m$ in the dataset. REMOVE TOO
In Fig.~\ref{fig:evolution}, the analysis illustrates how the models generalize, overfit, or underfit EMF exposure. 

To validate our RMSE results, Fig.~\ref{fig:evolution} - Row 1 illustrates that EME-CNTK$_{EigenPro}$ closely reflects the behavior of the input variables. This indicates that the model has successfully learned the inherent structure and characteristics of the exposure data. Consequently, the predictions closely match the underlying patterns and trends of the feature variables. Such fidelity to the original data dynamics highlights the effectiveness of CNTK in delivering accurate estimations of EMF exposure.

In contrast, GLIP tends to overfit the data, displaying poor generalization on the larger $128\times128$ grid. This is likely due to the inherent architecture of the GLIP CNN model. Adjusting its parameters could improve estimation accuracy. A similar pattern is observed in Fig.~\ref{fig:evolution} - row 3, where the CNTK model, trained with EigenPro, exhibits superior generalization compared to GLIP, further supporting the robustness of CNTK in larger grids. These results underscore the importance of model architecture and training methodologies in achieving accurate EMF exposure estimations.

\subsection{Comparative Analysis of Time Efficiency}

The efficiency of the proposed method was evaluated in terms of the time required for inference and training. Our analysis shows that the GLIP method required 73 seconds to train on 100 images for 150 epochs, on a machine with 12GB GPU memory (Intel core i7, NVIDIA RTX ADA 3500). In contrast, for 100 images, the EME-CNTK$_{EigenPro}$ approach took only 14 seconds to train for 350 epochs, i.e. 4.1.$10^{-4}$ seconds to train and impute one image as shown in Table \ref{tab:table4}.
% The table is given on the next page.
% \clearpage
\begin{table}[h]
\centering
\caption{Comparison of the proposed method with others with machine configuration and time}
\begin{tabular}{cccccc}
\toprule
Method & & Memory & Training  & Inference/ \\
& &usage &  & Test \\
\midrule
GLIP &  & 8GB RAM + & 73s & {-} \\ \vspace{0.1cm}
       &  & 380MB VRAM  \\
% Noisy-EMF: $U$ &  & 6GB RAM + &30 & {120s} & 1.5s \\ \vspace{0.1cm}
%         &  & 3.5GB VRAM \\
% Noisy-EMF: $N$ &  & 10GB RAM + &30& {17m 35s} & 0.5s \\ \vspace{0.1cm}
&  &  \\
EME-CNTK &  & 200MB RAM & 10s  & {3.5.$10^{-5}$ s} \\ \vspace{0.1cm}
 $exact$&  &  &   &  \\
EME-CNTK & & 300MB RAM & 14s & 4.1.$10^{-4}$ \\
 $EigenPro$&  & ~ 300MB VRAM &  &  \\ \bottomrule
\end{tabular}

\label{tab:table4}
\end{table}

Additional investigation is imperative to fine-tune sensor placement. The choice of the number of sensors to be deployed depends on the error that can be accepted and the cost of the deployment. Another constraint that needs to be studied in more detail (but is use-case dependent) is that the location of the sensors is limited to specific places, such as lampposts, and they must not be accessible to people. 

\section{Conclusion}\label{s4}
This study presents a comparative analysis of EMF exposure estimation from real-world sensors using both finite and infinite width convolutional networks. Notably, the EME-CNTK$_{EigenPro}$ model efficiently predicts accurate exposure maps with less than 1\% of the reference map area as input. Our findings reveal that infinite width networks achieve superior generalization on real-world data compared to the approach in \cite{9569470}. This is even true on higher resolution grids using a preconditioned gradient descent approach. However, further investigation is needed to enhance the prediction accuracy of GLIP, especially with low-density sensor networks. 
Future research should focus on improvement of the CNTK by adding skip connections, batch normalization and strided convolution in the infinite width CNN and identifying the specific sensor configurations that yield the most accurate EMF exposure estimates, balancing different performance metrics based on trade-offs.
\section*{Acknowledgment}
The work has been funded by the Métropole Européenne de Lille (MEL). The French government’s Beyond5G initiative, which was sponsored as a component of the country’s future investment program and strategy for economic recovery, provided also partial funding for this project. Discussions in the COST action CA20120 INTERACT are also invaluable source of inspiration. 
\bibliographystyle{ieeetr}

\bibliography{ref.bib}
\end{document}